\documentclass[11pt,a4paper]{article}
\usepackage{jcappub}

\usepackage{graphics}
\usepackage{epsfig}
\usepackage{graphicx}
\usepackage{bm}
\usepackage{subfig}

\newcommand{\ud}{\mathrm{d}}

\newcommand{\mbf}[1]{\mathbf{#1}}

\newcommand{\mosc}{m_{\mathrm{osc}}}

\newcommand{\tauosc}{\tau_{\mathrm{osc}}}
\newcommand{\chiosc}{\chi_{\mathrm{osc}}}

\def\beq{\begin{equation}}
\def\eeq{\end{equation}}
\def\baq{\begin{eqnarray}}
\def\eaq{\end{eqnarray}}
\def\k{{\bf k}}

\title{Non-Abelian dynamics in the resonant decay of the Higgs after inflation}

\author{Kari Enqvist,}
\author{Sami Nurmi}
\author{and Stanislav Rusak}

\affiliation{University of Helsinki and Helsinki Institute of Physics,\\P.O. Box
64, FI-00014, Helsinki, Finland }

\emailAdd{kari.enqvist@helsinki.fi}
\emailAdd{sami.nurmi@helsinki.fi}
\emailAdd{stanislav.rusak@helsinki.fi}

\abstract {We study the resonant decay of the Higgs condensate into
weak gauge bosons after inflation and estimate the corrections
arising from the non-Abelian self-interactions of the gauge fields.
We find that non-Abelian interaction terms induce an effective mass
which tends to shut down the resonance. For the broad resonance
relevant for the Standard Model Higgs the produced gauge particles
backreact on the dynamics of the Higgs condensate before the
non-Abelian terms grow large. The non-Abelian terms can however
significantly affect the final stages of the resonance after the
backreaction. In the narrow resonance regime, which may be important
for extensions of the Standard Model, the non-Abelian terms affect
already the linear stage and terminate the resonance before the
Higgs condensate is affected by the backreaction of decay products.}


\preprint{HIP-2014-06/TH}

\arxivnumber{1404.3631}

\begin{document}

\maketitle

\section{Introduction}

The measured Higgs mass in the range $M_{h}=125-126$ GeV
\cite{ATLAS:2012ae,Chatrchyan:2012tx} ensures vacuum stability
within the Standard Model (SM) up to very high energies -- orders of
magnitude above the electroweak scale. In particular, for top masses
sufficiently below the best fit value the SM remains stable up to
$\rho_{\rm inf}^{1/4}\sim 10^{16}$ GeV. Assuming the detection of
primordial gravitational waves by BICEP2 \cite{Ade:2014xna} will be
confirmed, this is the energy scale of inflation. If the Standard
Model is not significantly affected by new physics at these scales,
one generically predicts that the Higgs is a light field during
inflation ($m_{h}^2\ll H^2$) with a negligible contribution to the
total energy density \cite{DeSimone:2012qr,Enqvist:2013kaa}. As a
light field, during inflation the Standard Model Higgs acquires
long-wavelength fluctuations which in general render the local Higgs
value, averaged over the observable patch, different from the field
value at the minimum of the potential.

Small modifications of the SM Higgs potential, such as adding a
non-minimal Higgs-gravity coupling of the form $\xi H^{\dag}H R$,
are not expected to change the qualitative picture.  On the other
hand, it is well known that with an exceptionally large coupling,
($\xi \gg 1$), inflation might be realized with the Higgs field
itself \cite{Bezrukov:2007ep}. The predicted tensor-to-scalar ratio
of the Higgs inflation, $r=0.0033$ \cite{Bezrukov:2007ep}, is
however significantly below the reported detection of
$r=0.20^{+0.07}_{-0.05}$ by BICEP2, although the level $r\sim0.2$
could be achieved for a very specific choice of the SM parameters
(see e.g.
\cite{Bezrukov:2014bra,Masina:2014yga,Hamada:2014iga,Cook:2014dga}).

After inflation the observable patch of the universe is filled by an
effective Higgs condensate with small fluctuations around the
homogeneous background value. For the non-minimally coupled Higgs
inflaton the condensate dominates the energy density. If the
non-minial coupling is small (or zero), the energy density is
dominated by a non-Standard Model inflaton(s) (or its decay
products) while the Higgs condensate constitutes only a tiny
fraction of the total energy density. The Higgs fluctuations can
however be imprinted on metric perturbations even in this case if
the Higgs value affects the expansion history, for example through a
modulation of the inflaton decay rate
\cite{Dvali:2003em,Dvali:2003ar,Choi:2012cp}.

The dominant decay process of both the Standard Model Higgs and the
non-minimally coupled Higgs at zero temperature is the
non-perturbative production of weak gauge bosons from the
oscillating Higgs condensate \cite{Enqvist:2013kaa,Bezrukov:2008ut,
GarciaBellido:2008ab}. However, so far the non-Abelian
self-couplings of the SU$(2)$ gauge bosons have been neglected. In
investigating the resonant Higgs decay, the gauge fields have been
treated as if they were Abelian. While this arguably could be
justified at the onset of the resonance, the non-Abelian
interactions eventually become important as the number density of
the resonantly produced gauge field quanta grows large. This can
have a significant impact on the duration and efficiency of the
resonance.

In this work we investigate the non-perturbative decay of the
Standard Model Higgs into weak gauge bosons accounting for the
non-Abelian terms. Our primary goal is to estimate the time at which
the non-Abelian terms become significant for the dynamics of the
resonance. This should be compared with the time when the gauge
field induced mass for the Higgs, given by $m_{\rm ind}^2\sim g^2
\langle A^2\rangle$, becomes comparable to the effective mass
$\mosc^2\sim \lambda a^2h^2$ due to the Higgs amplitude $h$. The time
at which $m_{\rm ind}\gtrsim \mosc$ yields an estimate for the
beginning of backreaction of the resonantly produced gauge fields,
which eventually shuts down the resonance
\cite{Kofman:1997yn,Greene:1997fu}. If the non-Abelian terms were to
start to influence the gauge field dynamics when $m_{\rm ind}\ll
\mosc$, the Abelian approximation would fail already in describing
the linear stage of the resonance before the onset of backreaction.
We find that this is generically the case for the narrow resonance.
In the regime of broad resonance the Abelian approximation fails and
the non-Abelian terms grow large very soon after the linear stage of
the resonance. The non-Abelian terms will therefore play a
significant role in the subsequent non-linear regime after the
backreaction. The non-Abelian terms can therefore significantly
affect the decay of the Higgs condensate in the early universe.

The paper is organized as follows. In Section~\ref{sec:2} we outline
the initial conditions set by inflation and write down the relevant
equations. In Section~\ref{sec:3} we describe  resonant production
of weak gauge bosons from an oscillating Higgs condensate for the
Standard Model parameters and estimate the non-Abelian effects. In
Section~\ref{sec:4} we focus on the narrow resonance which requires
physics beyond the Standard Model. We conclude in
Section~\ref{sec:5} with a discussion of the results and their
implications.

\section{\label{sec:2}Equations of motion and initial conditions}

The tree-level potential of the Standard Model Higgs reads
  \beq
  V_0(h)=\lambda \left(h^2-\frac{\nu^2}{2}\right)^2 \ ,
  \eeq
where $h=|\Phi|$ is the magnitude of the complex Higgs doublet and
$\nu\simeq 246$ GeV. For large field values $h\gg \nu$ relevant for
the early universe the renormalization group improved effective
potential takes the form
  \beq
  V(h)\simeq \frac{\lambda(h)}{4}h^4\ .
  \eeq
In the $\overline{MS}$ renormalization scheme and for the best fit
SM parameter values the coupling becomes zero $\lambda(h_c)=0$ at
$h_c\sim 10^{10}$ GeV, indicating vacuum instability. The recent
measurement of the CMB B-modes by the BICEP2 collaboration
\cite{Ade:2014xna} has been claimed to suggests the existence of the background of
primordial gravitational waves with a tensor-to-scalar ratio $r\sim
\mathcal{O}(0.1)$, which implies an energy scale of inflation of
$H_* \sim 10^{14}$ GeV -- well above the instability scale for the
central values of the SM parameters. The Higgs effective potential
exhibits a maximum near the instability scale, and in order for us
to find ourselves in the current electroweak vacuum, inflationary
perturbations must not push the field over the maximum. This amounts
to a consistency condition $H_*<V_{\mathrm{max}}^{1/4}$
\cite{Enqvist:2014bua}. For the electroweak vacuum to remain stable up to
the inflationary scale, the values of the SM parameters need to be
at least $2\sigma$ away from their central values, or alternatively,
new physics must come into play at high energies to stabilize the
vacuum.

For field values below the instability scale the Higgs generically
is a light field. Provided that inflation lasts at least a few tens
of e-foldings longer that the $N_{\rm CMB}\sim 60$ e-foldings
corresponding to the presently observable patch, the Higgs
fluctuations on superhorizon scales will settle down to an
equilibrium state with the distribution given by $P(h)\sim
e^{-8\pi^2V/3H^4}$ \cite{Starobinsky:1986,Starobinsky:1994bd}. An
estimate of the typical value of the Higgs condensate over a patch
of the size of observable universe is then given by the root mean
square of the fluctuations
  \beq
 \label{eq:equilibrium}
  h_{*} = \sqrt{\langle h^2\rangle} \simeq 0.36 \lambda_{*}^{-1/4}
  H_{*}\ .
  \eeq
In the following we will use this estimate as the initial condition
for the evolution of the Higgs condensate after the end of
inflation.

After the end of inflation the Higgs condensate eventually starts to
oscillate and decays through a non-perturbative resonance into weak
gauge bosons \cite{Enqvist:2013kaa}. Here we shall assume that there
is not yet any thermal background due to the inflaton decay; in
effect, we assume that the inflaton decays slowly as compared with
the Higgs decay. In conformal time, the relevant part of the SM
action containing the SU$(2)\times$U$(1)$ gauge fields and the Higgs
sector reads
\beq
 S = -\int \ud^4x \left\{\frac{1}{4} \eta^{\mu\alpha}\eta^{\nu\beta}\Big(F_{\mu\nu}^a F_{\alpha\beta}^a + G_{\mu\nu}G_{\alpha\beta}\Big) +
 a^4\left[\left(D_{\mu}\Phi\right)^{\dagger}D^{\mu}\Phi + \frac{\lambda(\Phi^{\dag}\Phi)}{4}(\Phi^{\dag}\Phi)^2\right]
 \right\}\ ,
\eeq
where the kinetic terms are given by\footnote{In fact, in this case curved spacetime covariant derivatives $\nabla_{\mu}$ can be replaced by partial derivatives $\partial_{\mu}$.}
\baq
  F^a_{\mu\nu} &=& \nabla_{\mu}A_{\nu}^a - \nabla_{\nu}A_{\mu}^a +
  g\epsilon^{abc}A_{\mu}^bA_{\nu}^c\ , \\\nonumber
  G_{\mu\nu} &=&
  \nabla_{\mu}B_{\nu} - \nabla_{\nu}B_{\mu}\ ,\\\nonumber
  D_{\mu}\Phi &=& \left(\nabla_{\mu} -igA^a_{\mu}\tau^a -
  \frac{i}{2}g'B_{\mu}\right)\Phi\ .
\eaq
We choose to work in the unitary gauge where the three Goldstone
modes vanish and the Higgs doublet takes the form
  \beq
  \Phi = \frac{1}{\sqrt{2}}\left(\begin{array}{c} 0 \\
  \chi/a
  \end{array}\right) \ ,
  \eeq
where we have denoted the rescaled Higgs modulus by $\chi \equiv a
h$. The equation of motion for the Higgs then reads

\begin{eqnarray}
 \label{eq:higgseom}
 \ddot \chi + \left[-\frac{\nabla^2}{a^2} - \frac{\ddot a}{a} - \lambda\nu^2a^2 + \lambda \chi^2\right]\chi & = & -\frac{1}{4}\left[g^2A^a_{\mu}A^{a{\mu}} - 2gg'B_{\mu}A^{3\mu} + g'^2B_{\mu}B^{\mu}\right]\chi
\end{eqnarray}
where the dots denote derivatives with respect to conformal time
$d\tau \equiv a^{-1}dt$ and the indices are raised and lowered with the Minkowski metric $\eta_{\mu\nu}$. For spatial components of the gauge fields the
equation of motion is given by
\begin{eqnarray}
 \label{eq:eom_Aai}
 \ddot A^a_i - \nabla^2A^a_i - \partial_i(\dot A^a_0 - \partial_jA^a_j) + \frac{g^2\chi^2}{4}A^a_i & = & g\epsilon^{abc}\eta^{\mu\nu} \left[\partial_{\mu}(A^b_{\nu}A^c_i)
 + A^b_{\mu}\partial_{\nu}A^c_i - A^b_{\mu}\partial_iA^c_{\nu}\right] \\ &&  +\: g^2\eta^{\mu\nu}\left[A^a_{\mu}A^b_{\nu}A^b_i - (A^b_{\mu}A^b_{\nu})A^a_i\right] +
 \frac{gg'\chi^2}{4}\delta^{a3}B_i\ , \nonumber
\end{eqnarray}
and the $0$-components obey the constraint
\begin{eqnarray}
 \nabla^2A^a_0 - \partial_0(\partial_i A^a_i) - \frac{g^2\chi^2}{4}A_0^a & = & -g\epsilon^{abc}\left[\partial_i(A^b_iA^c_0) + A^b_i\partial_i A^c_0 - A^b_i\dot A^c_i\right]
 \\
 && - \:g^2\left[(\eta^{\alpha\mu}A^a_{\mu}A^b_{\alpha})A^b_0 - (\eta^{\alpha\mu}A^b_{\mu}A^b_{\alpha})A^a_0\right] -
 \frac{gg'\chi^2}{4}\delta^{a3}B_0\ . \label{eq:Aeom} \nonumber
\end{eqnarray}

The kinetic part of the gauge field action has conformal
symmetry. Therefore the mode functions of the gauge fields do not directly feel
the expansion of spacetime during inflation and, consequently,
the inflationary stage does not generate semiclassical
long-wavelength gauge field fluctuations. The initial conditions
for the gauge field dynamics in the post-inflationary epoch are
thus given by the vacuum solution.

Having specified the equation of motion and the initial
conditions for the system we now move on to study the dynamics and
the decay of the Higgs condensate.

\section{\label{sec:3}Non-perturbative decay of the Higgs condensate}

\subsection{Dynamics of the Higgs condensate}

Neglecting interactions of the Higgs condensate with other fields,
its equation of motion reads
\beq
 \ddot \chi + \lambda \chi^3 - \frac{\ddot a}{a}\chi = 0. \label{eq:chiEOM}
\eeq
For a constant equation of state, after inflation  the scale factor of the universe
evolves as $a =
\left[1+\frac{1}{2}(1+3w)H_*\tau\right]^{\frac{2}{1+3w}}$, where
$w=\rho/p$ and star denotes end of inflation. If the
inflaton oscillates in a quartic potential or is assumed to decay
instantaneously into radiation\footnote{However, if the inflaton decays
before the Higgs condensate starts to oscillate then the subsequent dynamics
will be affected by the thermal bath produced by the inflaton. In particular,
the fields will acquire thermal masses which tend to block resonant production
of particles \cite{Enqvist:2012tc,Enqvist:2013qba}.}, $w=1/3$, and the last term in
equation~\eqref{eq:chiEOM} vanishes. If the inflaton oscillates in a
harmonic potential, as is more commonly the case, the evolution is
matter-like ($w=0$) and $\chi$ grows initially until the last term
becomes subdominant. This happens at $\tauosc \sim H_*^{-1}$.
Afterwards the equation of motion is of the elliptic form and has a
solution that can be expressed in terms of a Jacobi elliptic cosine:

\begin{equation}
 \chi=\chiosc\operatorname{cn}\left[\sqrt{\lambda \chiosc^2}(\tau-\tauosc),\frac{1}{\sqrt{2}}\right].
\end{equation}

\subsection{Abelian dynamics}

Since the non-linear terms are small in the beginning, we treat the initial stage of particle production by considering the Abelian part of the equations of
motion. Particle production is caused by the oscillating mass
induced by the interaction with the Higgs condensate. Therefore we are interested in the mass eigenstates $W_{\mu}^{\pm}$
(alternatively $A_{\mu}^{1,2}$) and $Z_{\mu} = \cos\theta_W A^3_{\mu} - \sin\theta_W B_{\mu}$, where $\theta_W = \arctan \frac{g'}{g}$
is the weak mixing angle. The photon field $\mathcal A_{\mu} = \sin\theta_W A^3_{\mu} + \cos\theta_W B_{\mu}$
does not acquire a mass through the Higgs mechanism so that there is no resonant production of photons.

Vector fields may be decomposed into divergence-free
and curl-free parts, which in momentum space correspond to components that are transversal and longitudinal to
the momentum vector $\mathbf k$, satisfying respectively $\mathbf{k\cdot A}_{\mbf{k}}^T = 0$ and $\mathbf{k \times A}_{\mbf{k}}^L = 0$.
Defining new variables as

\begin{equation}\label{newpara}
  z \equiv \sqrt{\lambda \chiosc^2}(\tau-\tau_{\mathrm{osc}}), \qquad \kappa^2 \equiv \frac{k^2}{\lambda\chiosc^2},
  \qquad q \equiv \left\{\begin{array}{cl}\frac{g^2}{4\lambda} & \text{for the W bosons} \\ \frac{g^2+g'^2}{4\lambda} & \text{for the Z boson}\end{array}\right.,
\end{equation}
the transversal components obey the Lam\'{e} equation

\begin{equation}
 \label{eq:lame}
 \frac{\ud^2\mbf A^T}{\ud z^2} + \left[\kappa^2 + q\operatorname{cn}^2\left(z,\frac{1}{\sqrt{2}}\right)\right]\mbf A^T = 0.
\end{equation}
The setup is analogous to the case of massless preheating: the solutions get resonantly amplified within specific bands of momenta
and the structure of resonance has been studied in detail in~\cite{Greene:1997fu}. The longitudinal component obeys the equation

\begin{equation}
 \frac{\ud^2 A^L}{\ud z^2} + \frac{2\kappa^2}{\kappa^2 + q(\chi/\chiosc)^2}\frac{\ud\ln\chi}{\ud z}\frac{\ud A^L}{\ud z} + \left[\kappa^2 + q^2\operatorname{cn}^2\left(z,\frac{1}{\sqrt{2}}\right)\right] A^L  =  0.
\end{equation}
Compared to the transversal components the longitudinal one acquires a friction term resulting from the time dependence of the
effective mass induced by the interaction with the Higgs condensate. This prevents the resonance from occurring and the important
contribution to the dynamics will come from the transverse modes. Thus in what follows we shall concentrate on the transverse modes alone.

\begin{figure}
\centering
\includegraphics[width=0.45\textwidth]{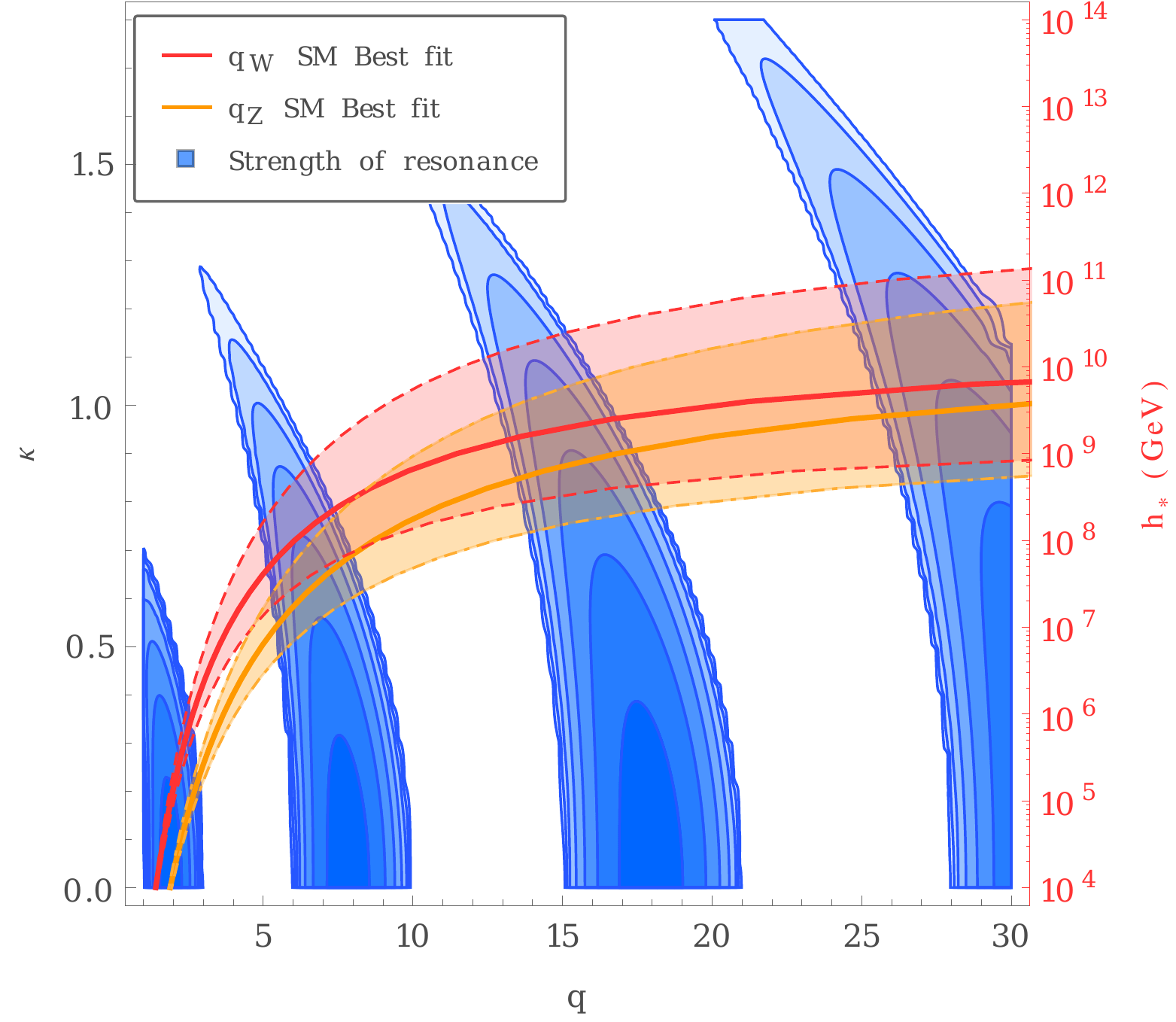}
\caption{\label{fig:structure_of_resonance}Strucure of resonance for the Lam\'{e} equation~\eqref{eq:lame}. Solutions are exponentially amplified within the
blue bands. Also plotted the dependence of the resonance parameter $q$ on energy scale (initial value of the Higgs) for the $W$ (red) and $Z$ (orange) bosons. The curves correspond
to central values $M_{\mathrm{top}} = 173.1$ GeV, $M_{\mathrm{Higgs}} = 125.66$ GeV and $\alpha_s = 0.1184$ while the shaded regions correspond to $1\sigma$ uncertainty in the top mass.}
\end{figure}

In principle, the dynamics of the resonance are fully specified by
the couplings of the Standard Model since the strength of
amplification is determined only by their combination $q$
(\ref{newpara}). However, since the couplings run with energy, their
values will depend on the energy scale of inflation\footnote{The
amplitude of the Higgs condensate in our local patch could also
happen to be different from the equilibrium root mean square
value~\eqref{eq:equilibrium}, in which case the energy scale
determining the effective couplings would not be directly related to
the scale of inflation.}. Figure~\ref{fig:structure_of_resonance}
shows the structure of resonance together with the energy dependence
of the resonance parameter $q$ for weak gauge bosons. With the
inflationary scale fixed by the amplitude of gravitational waves the
dynamics are completely determined and taking into account the
running of the couplings at two-loop
order~\cite{Degrassi:2012ry,Buttazzo:2013uya}, the system is in the
intermediate to broad resonance regime, with $q = 1..100$
 unless we are extremely close to the instability scale.
The energy scale at the end of inflation will
determine the values of the couplings and by extension the strength
of the resonant production, as well as the location of the resonance
bands in momentum space. Note that the broader end of the resonance
is sensitive to the uncertainties in the measurements for the masses
of the Higgs and, in particular, the top quark. The Higgs also produces its own
quanta; however, this process is comparatively inefficient because it is located
right on the edge of a resonance band with $q_h=3$.

One might also be concerned that the produced gauge bosons would decay perturbatively into fermions before the resonance has a chance to build on them.
This will happen if their decay rate is larger than the frequency of the Higgs oscillations. The time between two zero-crossings of the Higgs is given
by $\Delta \tau = 2K\mosc^{-1}$, where $K$ is the complete elliptic integral of the first kind, while the total decay rate of the gauge bosons is $\Gamma_Z \simeq \frac{2}{3\cos^3\theta_W}\Gamma_{W^{\pm}} \simeq \frac{g^2 m_Z}{8\pi\cos^3\theta_W}=
\frac{g^2\sqrt{q_Z}\mosc}{8\pi\cos^2\theta_W}$ \cite{GarciaBellido:2008ab}. Therefore, they will decay in the time interval between zero-crossings if

\begin{equation}
 q > \left(\frac{4\pi\cos^2\theta_W}{g^2K}\right)^{2} \sim \mathcal O(10^2).
\end{equation}
Thus, the decay of the produced gauge bosons is only relevant if the
Higgs self-coupling is close to the instability scale and we ignore it in our analysis.

For definiteness, in the following we take the energy scale to be
$H_*=10^{14}$ GeV, as implied by BICEP2 \cite{Ade:2014xna}, and
choose $M_{h} = 125.7$ GeV, $M_t = 171.25$ GeV and $\alpha_s =
.1191$ ($\lesssim 2\sigma$ away from central values) to ensure
vacuum stability. The resonance parameters are then $q_W \simeq 18$
and $q_Z \simeq 29$. Note that typically either $W$ or $Z$ will be
dominantly produced depending on which one lies within the strongest
resonance band for a given energy scale and/or choice of parameters.

\subsection{Non-Abelian corrections}

As the modes get amplified, the non-linear interaction terms in equations \eqref{eq:eom_Aai} grow to a size comparable to the linear terms. We use the Hartree
approximation to estimate the effect of these terms on the dynamics of the resonance, i.e., we take the vacuum expectation value of the non-linear terms
using the linear solution

\begin{equation}
 A^{(W,Z)}_i(\mbf x) = \int \frac{\ud^3 \mbf k}{(2\pi)^{3/2}}\sum_{\lambda=1}^3\left[A^{(W,Z)}_{\mbf k,\lambda}\epsilon_i(\mbf k,\lambda)\hat a^{(W,Z)}_{\mbf k,\lambda}e^{i\mbf{k\cdot x}}
                                                                                    + A^{(W,Z)*}_{\mbf k,\lambda}\epsilon_i^*(\mbf k,\lambda)\hat a^{(W,Z)\dagger}_{\mbf k,\lambda}e^{-i\mbf{k\cdot x}}\right],
\end{equation}
where $\hat{a}_{\k},\hat{a}^{\dag}_{\k}$ are the usual annihilation and creation operators, $\epsilon_i(\mbf k,\lambda)$ are the polarization vectors
and $A^{(W,Z)}_{\mbf k,\lambda}$ are the solutions to the corresponding equations of motion.
We choose the vector $\epsilon_i(\mbf k,3) = k_i/|\mbf k|$ corresponding to the longitudinal component and the remaining two vectors to be orthogonal to
$\mbf k$, corresponding to transversal components, and satisfying

\begin{equation}
 \sum_{\lambda=1}^2 \epsilon_i(\mbf{k},\lambda)\epsilon_j^*(\mbf{k},\lambda) = \delta_{ij} - \frac{k_ik_j}{k^2}.
\end{equation}
The usual quantization condition for the fields and their conjugate momenta then also imposes the Wronskian condition $A\dot A^* - A^*\dot A=i$.
For the initial amplitudes we take the WKB solution $A = \frac{1}{\sqrt{2\omega}}e^{-i\int \omega_k\ud \tau}$ at $\tauosc$.

Since the fields are treated as independent the antisymmetric interaction terms vanish in the Hartree approximation and the only contribution comes
from the trilinear terms so that the right hand side of equation~\eqref{eq:eom_Aai} for the component $A^a_i$ will then give

\begin{equation}
  -g^2\eta^{\mu\nu}\sum_{b \neq a}\left[\left<A^b_{\mu}A^b_{\nu}\right>A^a_i - \left<A^b_{i}A^b_{\mu}\right>A^a_{\nu}\right].
\end{equation}
The temporal component is tied to the longitudinal one and so is likewise not amplified. The spatial two-point function for the mass eigenstates is given by

\begin{figure}
\subfloat{
\includegraphics[width=0.45\textwidth]{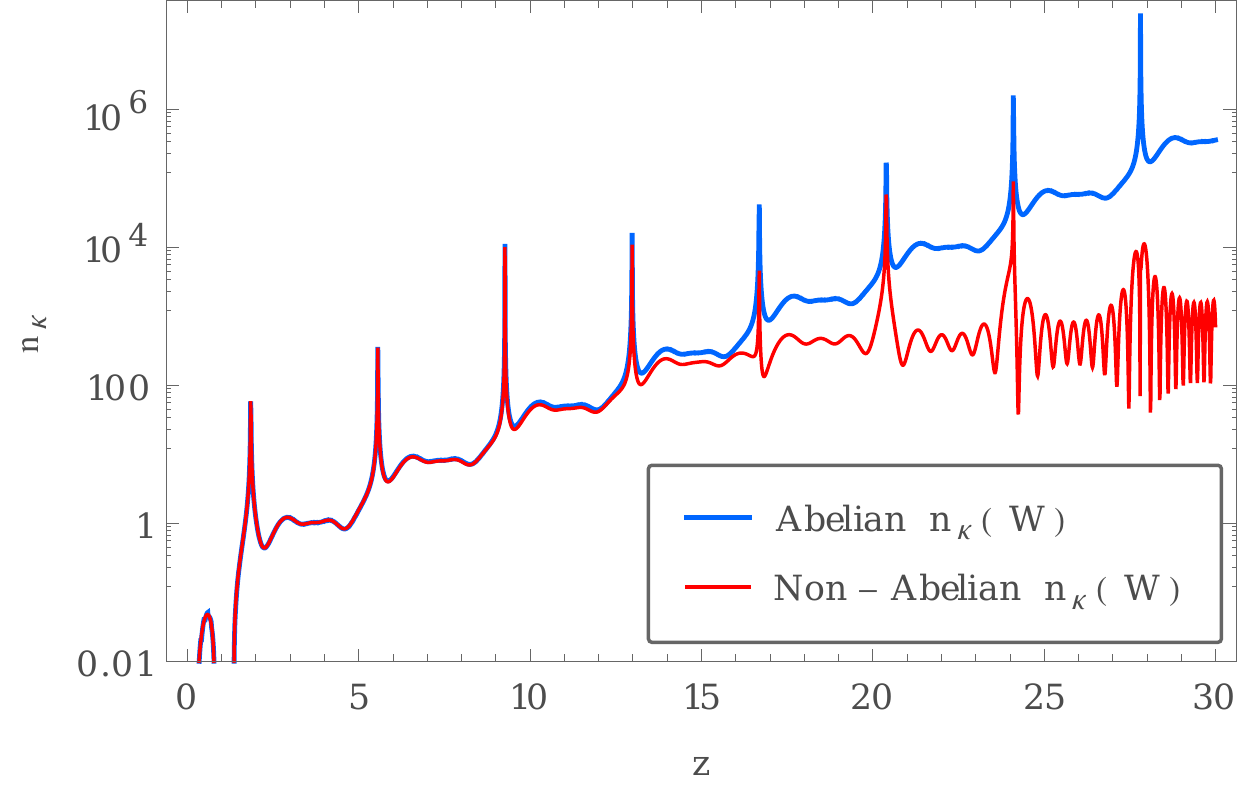}}
\quad
\subfloat{
\includegraphics[width=0.45\textwidth]{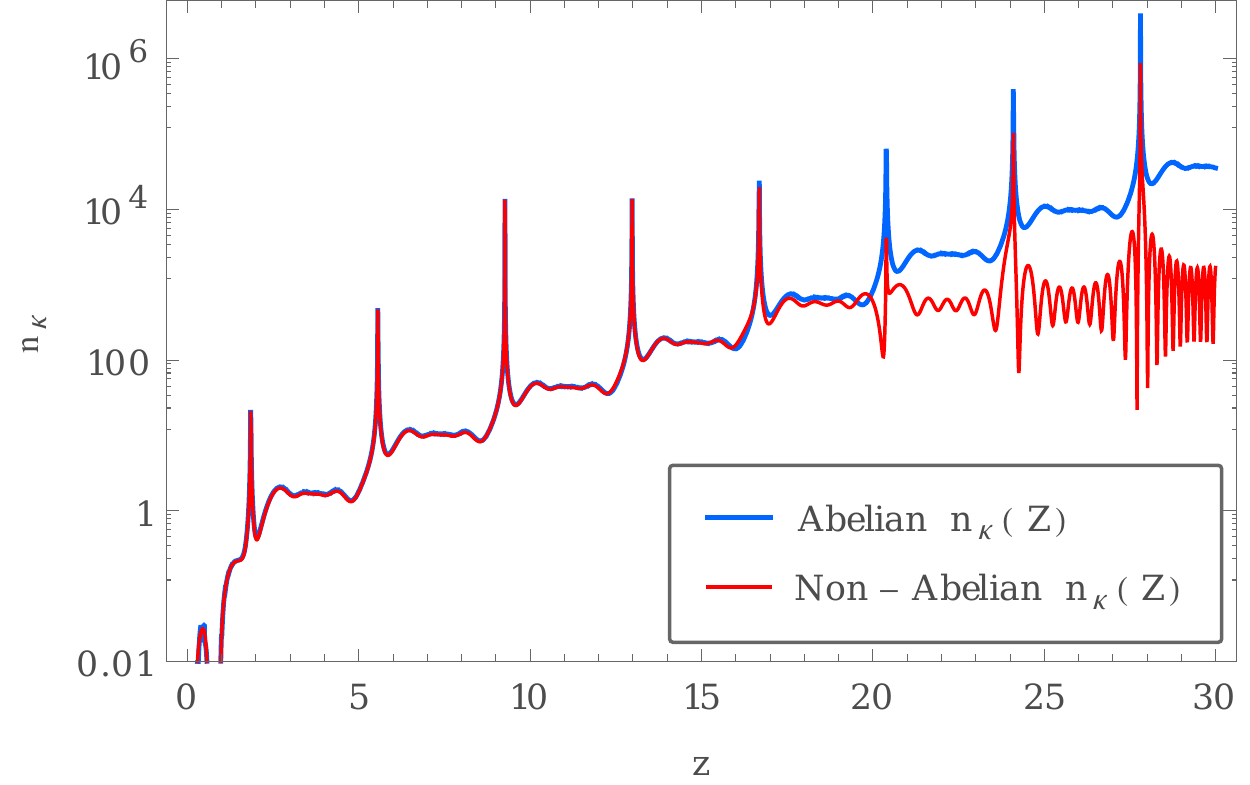}}
\caption{\label{fig:termination_of_resonance}Comparison of particle densities $n_{k=0}$  between Abelian dynamics
and the case where the non-Abelian mass term has been taken into account. The induced mass terminates the resonance once it becomes comparable to the mass
coming from the Higgs. Here $z$ is the scaled time defined in Eq. (\ref{newpara}).}
\end{figure}

\begin{eqnarray}
  \left< A_i^{(W,Z)}A_j^{(W,Z)}\right> 
%
& = & \frac{\delta_{ij}}{3\pi^2} \int_0^{\infty}\ud k\:k^2\left(\left|A_{(W,Z)}^T\right|^2 + \frac{1}{2}\left|A_{(W,Z)}^L\right|^2\right) \simeq \frac{\delta_{ij}}{3\pi^2} \int_0^{\infty}\ud k\:k^2\left|A_{(W,Z)}^T\right|^2
\nonumber
\end{eqnarray}
The transversal components therefore acquire an effective mass term induced by the non-linear interactions. The induced masses for $W$ and $Z$ bosons are respectively

\begin{eqnarray}
 m_W^2 & = & \frac{2g^2\lambda \chiosc^2}{3\pi^2}\int_0^{\infty}\ud \kappa \kappa^2\Big(|X_W|^2 + \cos\theta_W^2|X_Z|^2\Big), \\
 m_Z^2 & = & \frac{4g^2\lambda \chiosc^2 \cos\theta_W}{3\pi^2}\int_0^{\infty}\ud \kappa \kappa^2|X_W|^2,
\end{eqnarray}
where $X \equiv (\lambda\chiosc^2)^{1/4} A$ are independent of initial field values. The integrals are in principle infinite and need to be regularized
but we are only interested in the particles produced by the resonance, which are in the IR regime, so we can integrate up to the cutoff $\kappa \sim q^{1/4}$ \cite{Kofman:1997yn,Greene:1997fu}. This induced mass terminates the resonance once it
becomes comparable to the mass coming from the interaction with the Higgs condensate. Figure~\ref{fig:termination_of_resonance} compares the number of produced particles $n_{k}$ in
the Abelian case and the case with a mass induced by non-Abelian terms, where

\begin{equation}
  n_{k} \equiv \frac{\omega_k}{2}\left[\frac{|\dot{A}|^2}{\omega_k^2} + |A|^2\right] - \frac{1}{2}.
\end{equation}
Note that the $Z$ boson contributes only to the effective mass of the $W$ bosons but not to its own mass. Therefore, in principle,
if the charged weak bosons are in a less efficient resonance band then the produced $Z$ particles will shut down the $W$
production before it can affect the production of $Z$ particles leaving the latter unimpeded. However, we expect higher order loop
effects, which are not captured by the Hartree approximation, to come into play at that point. In that case a more detailed
analysis using lattice simulations is needed and we leave this for a future publication. We concentrate instead on the case
where $W$ production is more efficient.

\subsection{Back-reaction to the Higgs}

\begin{figure}
\centering
\includegraphics[width=0.45\textwidth]{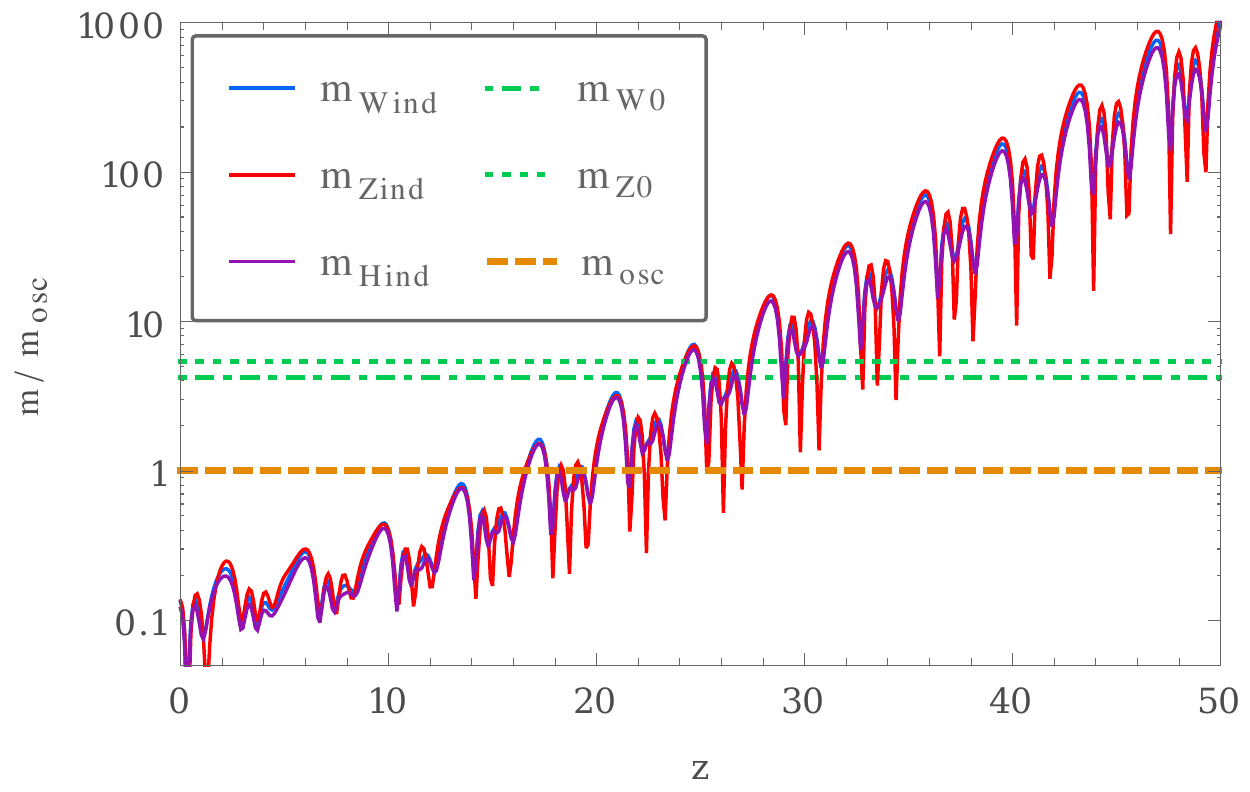}
\caption{\label{fig:mass_comparison}Comparison between the
different contributions to the effective masses. Blue
and red curves show the effective masses induced by non-Abelian
interactions for W and Z bosons respectively. The green lines
correspond to the initial masses contributed from interaction with
the Higgs condensate for W (dot-dashed) and Z (dotted). The purple line shows the contribution to
the effective Higgs mass from produced gauge bosons while the dashed
orange line is the original mass of the Higgs condensate $\mosc \equiv
\sqrt{\lambda\chiosc^2}$.}
\end{figure}

The interactions among the weak gauge bosons also induce an effective mass for the Higgs condensate. From equation~\eqref{eq:higgseom} this is given by

\begin{equation}
 m_{\chi}^2 = \frac{g^2}{4}\Big[2\langle W^+_{\mu}W^{\mu-}\rangle + (\cos\theta_W)^{-2}\langle Z_{\mu}Z^{\mu}\rangle\Big].
\end{equation}
Since the induced mass for the gauge bosons is $m_{W,Z}^2 \sim q
m_{\chi}^2$,  in the broad resonance regime the correction to the
Higgs equation of motion will become important earlier than the
correction to the dynamics of gauge bosons themselves. However,
since the resonance is not extremely broad for the Standard Model
parameters, with a typical resonance parameter given by $q \sim
\mathcal{O}(10)$, the difference is rather small and the non-Ablian
corrections may still require attention even during the final stages
of the linear regime of the resonance. Of course, this very much
depends on the actual values one adopts for the SM parameters.
Figure~\ref{fig:mass_comparison} shows the comparison between
different contributions to the effective masses of the fields.

Another effect that may be relevant for the dynamics of the resonance is the interaction
of the produced gauge bosons with the oscillating field $\chi$, often 
referred to as `rescattering'. The oscillating field $\chi$ may be thought
of as a condensate of particles at rest. A crude estimate for the importance
of rescattering is given by \cite{Kofman:1997yn}, using the usual methods of scattering between
$\chi$ and gauge boson particles. The time for each
boson to scatter once with a particle in the condensate is

\begin{equation}
 \tau > \sqrt{q}\mosc^{-1}\left(\frac{\mosc}{m_{\chi}}\right)^2
\end{equation}
where $m_{\chi}$ is the effective mass of $\chi$ induced by the produced gauge
bosons. When backreaction is negligible and $q \sim \mathcal O(10)$ this appears to be much larger than the oscillation
time $\mosc^{-1}$ and rescattering becomes important only after backreaction sets in.
Therefore rescattering does not appear to significantly affect our estimates for the onset of backreaction.

While it seems justified to neglect the non-Abelian terms as a first
approximation under the linear stage of the resonance their impact
is essential when considering the non-linear regime after the
back-reaction. We plan to address the dynamics after the
back-reaction in a future work using a full lattice simulation. A
detailed investigation of the resonance dynamics is indeed required
to determine the time when the Higgs condensate eventually has fully
decayed.

\section{\label{sec:4}Narrow resonance}

As we have shown above, for a broad resonance, the produced gauge bosons backreact on the Higgs before their own evolution dynamics is significantly affected. Here we consider
the narrow resonance regime where the opposite effect occurs. Narrow resonance may be relevant if some physics beyond the Standard Model affect
the running of the couplings or if the decaying field is other than the Higgs. In these cases
the couplings are no longer determined by the SM beta functions so that we may choose arbitrary values. For simplicity, let us consider $g'\ll g$ so that $q_Z \simeq q_W$ and $\cos\theta_W\simeq 1$.

\begin{figure}[t!]
\subfloat[][]{
\includegraphics[width=0.45\textwidth]{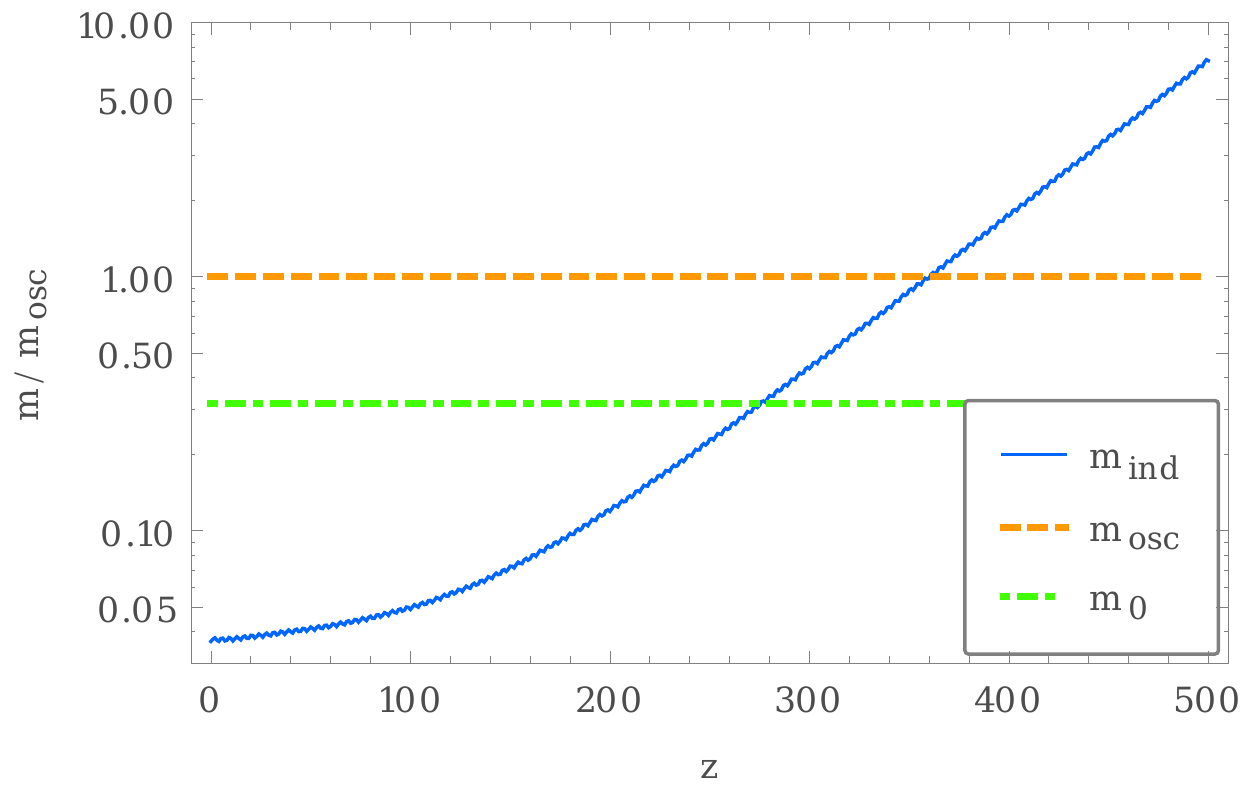}}
\quad
\subfloat[][]{
\includegraphics[width=0.45\textwidth]{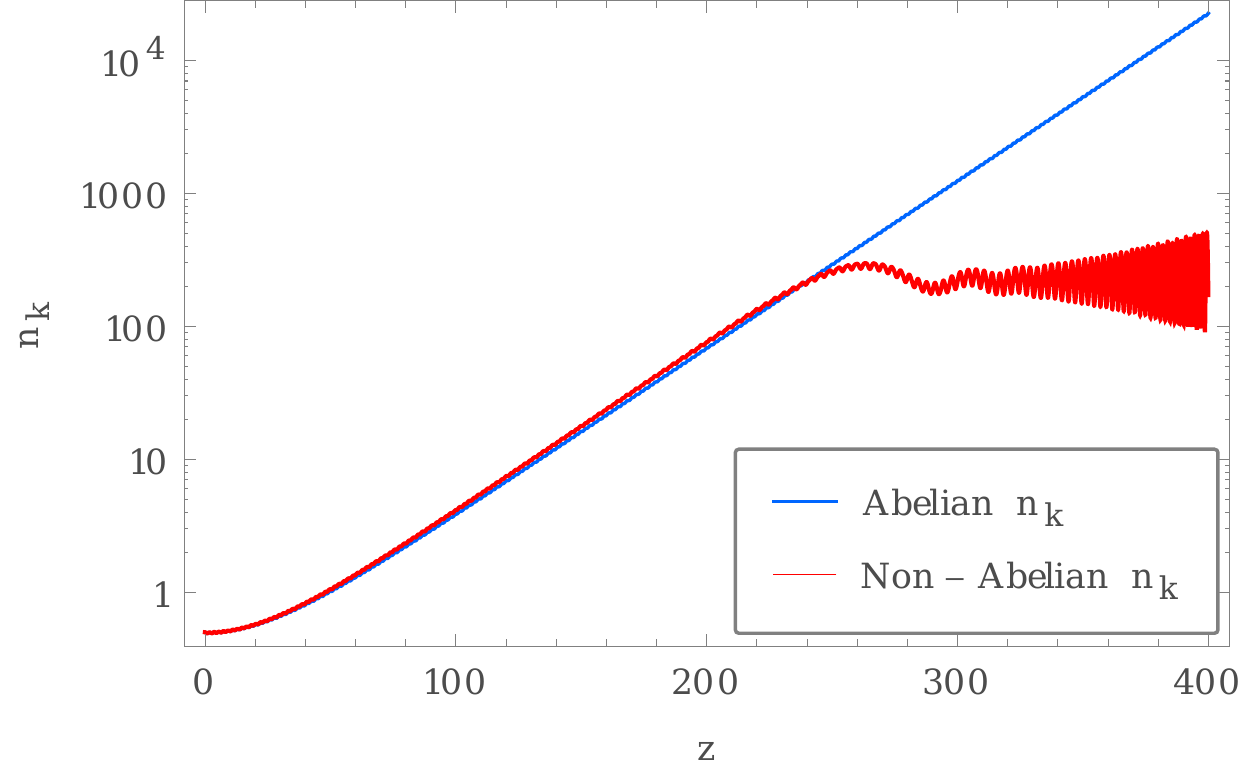}}
\caption{\label{fig:narrow_resonance}Left: mass induced by non-Abelian interactions for the narrow resonance case $q = 0.1$ (blue solid line).
The green dot-dashed line is the original gauge boson mass due to interaction with the Higgs $m_0$ and the dashed orange line is the
original mass of the Higgs condensate $\mosc \equiv \sqrt{\lambda\chiosc^2}$.
Right: termination of the production of particles for $\kappa = \kappa_{\mathrm{center}}$ by the non-Abelian effective mass.}
\end{figure}

Approximating $\operatorname{cn}\left(z,1/\sqrt{2}\right) \simeq \cos\left[\frac{\pi z}{2K}\right]$, where $K$ is the complete
elliptic integral of the first kind, and defining

\begin{equation}
 \tilde\kappa^2 \equiv \frac{4K^2}{\pi^2}\left(\kappa^2 + \frac{q}{2}\right) \qquad \tilde q \equiv \frac{K^2}{\pi^2}q \qquad x \equiv \left(\frac{1}{2}+\frac{z}{2K}\right)\pi
\end{equation}
we can write the equations of motion as the Mathieu equation

\begin{equation}
 X'' + \left(\tilde\kappa^2 -2\tilde q\cos 2x \right)X = 0.
\end{equation}
Solutions are known to get amplified in a narrow bands $\tilde \kappa^2 \sim l^2 \pm \tilde q^l$, where $l = 1,2,3,...,$ with the dominant contribution coming from the first such band \cite{Kofman:1997yn}.
The center of the band is at

\begin{equation}
 \kappa_{\mathrm{center}}^2 = \frac{\pi^2}{4K} - \frac{q}{2}.
\end{equation}
The mass induced by the production of particles shifts the location
of the band as $\kappa^2 \rightarrow \kappa^2 +
\frac{m_{\mathrm{ind}}^2}{\mosc^2}$ so as the induced mass grows,
the resonance band moves towards the origin until it reaches it
after which no modes are excited any more. This happens when
$\frac{m_{\mathrm{ind}}^2}{\mosc^2} \sim \frac{\pi^2}{4K}$ which
holds true when $m_{\mathrm{ind}}$ is of order $\mosc \equiv
\sqrt{\lambda\chiosc^2}$. However, the resonance becomes inefficient
even before that because the width of the resonance band is $\sim q$
so that the band moves away from its original location when the
induced mass reaches $\sim \sqrt{q}\mosc$ (i.e. the initial gauge
boson mass). After that the resonance can no longer build on the
particles that it has been producing until then.
Figure~\ref{fig:narrow_resonance} shows the growth of the effective
mass induced by non-Abelian terms and the resulting termination of
particle production. As in the broad resonance regime, the effective
mass induced by non-Abelian interactions shuts down the resonance,
but in contrast to broad resonance this happens before the dynamics
of the Higgs are affected. Therefore, non-Abelian terms should be
taken into account in the narrow resonance regime already in
investigating the dynamics before the onset of back-reaction.

\section{\label{sec:5}Conclusions}

We have studied the effects of non-Abelian interactions in the
resonant decay of the Higgs condensate after inflation. We found
that the self-interactions of the weak gauge fields induce an
effective mass term which grows quickly and shuts down the
resonance. However, for the Standard Model parameters, the
non-perturbative decay of the Higgs condensate is in the
intermediate to broad regime, and in this case the gauge fields
backreact on the Higgs dynamics before their non-Abelian terms will
come to dominate their own equations of motion. This suggests that,
as a first approximation, the non-Abelian terms can be justifiably
neglected before the backreaction and the system can be approximated
by an Abelian one. Nevertheless, since the resonance is only
moderately broad for the Standard Model couplings (unless the energy
is very close to the instability scale), the time difference between
the produced particles backreacting on the Higgs dynamics and on the
gauge fields themselves is rather short. Therefore, the effect of
non-Abelian terms may still play a role during the final stages of
the linear regime before the backreaction. Moreover, the non-Abelian
terms appear to play a significant role during the non-linear stage
after the backreaction which eventually determines the duration and
efficiency of the resonance.

We have also considered the case of narrow resonance, which can be
relevant if physics beyond the Standard Model intervenes at high
energies and changes the running of the couplings. This case is also
pertinent if the decaying field is taken to be something other than
the Higgs. For narrow resonance non-Abelian terms become important
before the Higgs dynamics are significantly affected and therefore
they need to be taken into account already in the linear regime
before the backreaction. The effect of the non-Abelian terms is to
shut down the resonance by shifting the resonance band in momentum
space so that the resonance can no longer build on previously
produced particles.

Since inflation inevitably generates a Higgs condensate, unless the
Standard Model Higgs sector is significantly modified, a detailed
understanding of the Higgs decay is an integral part in completing
the knowledge of the hot big bang epoch after inflation. The decay
of the Higgs condensate after inflation can have significant
consequences for subsequent physics, such as phase transitions
either within the SM or its modifications. Details of the Higgs
decay are also highly important for scenarios where the Higgs
sources curvature perturbations for example by modulating the decay
of the inflaton \cite{DeSimone:2012qr,Choi:2012cp}. In such
scenarios the Higgs should survive at least until the decay of the
inflaton and termination of the resonance by non-Abelian
interactions could facilitate this. Non-perturbative decay of the
Higgs condensate, as well as other spectator fields, would also
generate gravitational waves whose spectral features depend on
structure of the decay~\cite{Figueroa:2014aya}.

We have considered the non-Abelian corrections in the Hartree
approximation. A more detailed analysis using lattice simulations is
needed in order to gain a detailed understanding of the decay
process and in particular to address the dynamics after the
backreaction. We leave this for a future work. We have assumed the
inflaton decays slowly compared to Higgs and therefore we have not
considered possible thermal effects arising from the thermal bath
produced by the inflaton decay. If present, the buildup of the
thermal bath typically eventually blocks the resonance by inducing
large thermal masses for the
fields~\cite{Enqvist:2012tc,Enqvist:2013qba}. Other thermal effects,
such as dissipation of the field due to scattering off of particles in the
thermal bath may also become important~\cite{Mukaida:2012qn,Mukaida:2012bz}. However, previous
studies have focused on scalar fields alone, and it might be
interesting to consider thermal effects also within the context of
resonant production of non-Abelian gauge fields.

\acknowledgments

KE is supported by the Academy of Finland  grants
1263714 and 1263714, SN is supported
by the Academy of Finland grant 257532, and SR has been supported by the
V\"{a}is\"{a}l\"{a} and the Wihuri foundations.

\bibliographystyle{JHEP}
\bibliography{HiggsBib.bib}

\end{document}